\documentstyle[epsfig,12pt]{article}
\input{psfig.sty}
\newlength{\dinwidth}
\newlength{\dinmargin}
\newlength{\extraspace}
\newlength{\extraspaces}
\newcommand{\be}{\begin{equation}
\addtolength{\abovedisplayskip}{\extraspaces}
\addtolength{\belowdisplayskip}{\extraspaces}
\addtolength{\abovedisplayshortskip}{\extraspace}
\addtolength{\belowdisplayshortskip}{\extraspace}}
\newcommand{\ee}{\end{equation}}
\newcommand{\bdm}{\begin{displaymath}
\addtolength{\abovedisplayskip}{\extraspaces}
\addtolength{\belowdisplayskip}{\extraspaces}
\addtolength{\abovedisplayshortskip}{\extraspace}
\addtolength{\belowdisplayshortskip}{\extraspace}}
\newcommand{\edm}{\end{displaymath}}
\def\simlt{\mathrel{\lower2.5pt\vbox{\lineskip=0pt\baselineskip=0pt
           \hbox{$<$}\hbox{$\sim$}}}}
\def\simgt{\mathrel{\lower2.5pt\vbox{\lineskip=0pt\baselineskip=0pt
           \hbox{$>$}\hbox{$\sim$}}}}
\newcommand{\ls}[1]
   {\dimen0=\fontdimen6\the\font
    \lineskip=#1\dimen0
    \advance\lineskip.5\fontdimen5\the\font
    \advance\lineskip-\dimen0
    \lineskiplimit=.9\lineskip
    \baselineskip=\lineskip
    \advance\baselineskip\dimen0
    \normallineskip\lineskip
    \normallineskiplimit\lineskiplimit
    \normalbaselineskip\baselineskip
    \ignorespaces}

\catcode`@=11
\newcount\@tempcntc
\def\@citex[#1]#2{\if@filesw\immediate\write\@auxout{\string\citation{#2}}\fi
  \@tempcnta\z@\@tempcntb\m@ne\def\@citea{}\@cite{\@for\@citeb:=#2\do
    {\@ifundefined
       {b@\@citeb}{\@citeo\@tempcntb\m@ne\@citea\def\@citea{,}{\bf ?}\@warning
       {Citation `\@citeb' on page \thepage \space undefined}}%
    {\setbox\z@\hbox{\global\@tempcntc0\csname b@\@citeb\endcsname\relax}%
     \ifnum\@tempcntc=\z@ \@citeo\@tempcntb\m@ne
       \@citea\def\@citea{,}\hbox{\csname b@\@citeb\endcsname}%
     \else
      \advance\@tempcntb\@ne
      \ifnum\@tempcntb=\@tempcntc
      \else\advance\@tempcntb\m@ne\@citeo
      \@tempcnta\@tempcntc\@tempcntb\@tempcntc\fi\fi}}\@citeo}{#1}}
\def\@citeo{\ifnum\@tempcnta>\@tempcntb\else\@citea\def\@citea{,}%
  \ifnum\@tempcnta=\@tempcntb\the\@tempcnta\else
   {\advance\@tempcnta\@ne\ifnum\@tempcnta=\@tempcntb \else \def\@citea{--}\fi
    \advance\@tempcnta\m@ne\the\@tempcnta\@citea\the\@tempcntb}\fi\fi}
\catcode`@=12

\def\simge{\mathrel{%
   \rlap{\raise 0.511ex \hbox{$>$}}{\lower 0.511ex \hbox{$\sim$}}}}
\def\simle{\mathrel{
   \rlap{\raise 0.511ex \hbox{$<$}}{\lower 0.511ex \hbox{$\sim$}}}}
 
\def\slashchar#1{\setbox0=\hbox{$#1$}           
   \dimen0=\wd0                                 
   \setbox1=\hbox{/} \dimen1=\wd1               
   \ifdim\dimen0>\dimen1                        
      \rlap{\hbox to \dimen0{\hfil/\hfil}}      
      #1                                        
   \else                                        
      \rlap{\hbox to \dimen1{\hfil$#1$\hfil}}   
      /                                         
   \fi}                                         %
\def\nn{\nonumber}
\def\ts{\thinspace}

\def\ra{\rightarrow}
\def\Ra{\Rightarrow}

\def\ol{\bar}

\def\be{\begin{equation}} 
\def\ee{\end{equation}} 
\def\bea{\begin{eqnarray}}
\def\eea{\end{eqnarray}}
\def\ba{\begin{array}}
\def\ea{\end{array}}

\def\CH{{\cal H}}

\def\CM{{\cal M}}

\def\CO{{\cal O}}

\def\chv{\CH_{V_8}}
\def\chz{\CH_{Z'}}

\def\uone{U(1)_1}
\def\utwo{U(1)_2}
\def\uy{U(1)_Y}
\def\suone{SU(3)_1}
\def\sutwo{SU(3)_2}

\def\kslash{\raise.15ex\hbox{/}\kern-.57em k}

\def\mev{{\rm MeV}}
\def\gev{{\rm GeV}}
\def\tev{{\rm TeV}}

\def\half{{\textstyle{ { 1\over { 2 } }}}}

\def\thc{\theta_C}
\def\thy{\theta_Y}

\begin{document}
\title{
\vskip -15mm
\begin{flushright}
\vskip -15mm
{\small BUHEP-00-25\\
hep-ph/0012073\\}
\vskip 5mm
\end{flushright}
{\Large{\bf \hskip 0.38truein
$\ol B$--$B$ Mixing Constrains Topcolor--Assisted Technicolor}}\\
}
\author{
\centerline{{\small Gustavo Burdman\thanks{burdman@bu.edu}\ts\ts,}
{\small  Kenneth Lane\thanks{lane@physics.bu.edu}\ts\ts, and}
{\small Tongu\c c Rador\thanks{rador@buphy.bu.edu}}}\\
\centerline{{\small Department of Physics, Boston University,}}\\
\centerline{{\small 590 Commonwealth Avenue, Boston, MA 02215}}\\
}
\maketitle
\begin{abstract} 
We argue that extended technicolor augmented with topcolor requires that all
mixing between the third and the first two quark generations resides in the
mixing matrix of left--handed down quarks. Then, the $\ol B_d$--$B_d$ mixing
that occurs in topcolor models constrains the coloron and $Z'$ boson masses
to be greater than about 5~TeV. This implies fine tuning of the topcolor
couplings to better than 1\%.
\end{abstract}


\newpage

The impressive agreement of the standard model's predictions with
experimental data does not lessen the need for new physics to explain the
dynamics underlying electroweak and flavor symmetry breaking. This physics
may manifest itself not only in high energy collider experiments but also in
precision low energy measurements of meson decays and mixing. In turn, low
energy measurements powerfully constrain flavor physics scenarios. A prime
example is technicolor~\cite{tc} with extended technicolor
(ETC)~\cite{etc,tcreview}, a natural, dynamical scheme for electroweak and
flavor symmetry breaking. There, $\vert \Delta S\vert = 2$ effects in the
neutral kaon system require ETC gauge boson masses of $10^2$--$10^4\,\tev$
{\it and} walking technicolor to produce the correct first and second
generation quark masses. With such large masses, ETC by itself cannot account
for the top quark's mass. Therefore, we take it to be augmented by topcolor,
a system referred to as topcolor--assisted technicolor
(TC2)~\cite{topcref,tctwohill,tctwoklee}.

All TC2 models assume that color $SU(3)_C$ and weak hypercharge $\uy$ arise
from the breakdown of the topcolor groups $\suone\otimes\sutwo$ and
$\uone\otimes\utwo$ to their diagonal subgroups. Here $\suone$ and $\uone$
are strongly--coupled, $\sutwo$ and $\utwo$ are weakly--coupled, with the
color and weak hypercharge couplings given by $g_C = g_1 g_2 /\sqrt{g_1^2 +
g_2^2} \equiv g_1 g_2/g_{V_8} \simeq g_2$ and $g_Y = g'_1 g'_2/
\sqrt{g_1^{\prime \ts 2} + g_2^{\prime\ts 2}} \equiv g'_1 g'_2/g_{Z'} \simeq
g'_2$. Top and bottom quarks are $\suone$ triplets. The broken topcolor
interactions are mediated by a color octet of colorons, $V_8$, and a color
singlet $Z'$ boson, respectively. By virtue of the different $\uone$
couplings of $t_R$ and $b_R$, $V_8$ and $Z'$ exchange between third
generation quarks generates a large contribution $\hat m_t(1\,\tev) \simeq
160\,\gev$ to the top mass, but none to the bottom mass.

If topcolor is to provide a natural explanation of $\hat m_t$, the $V_8$ and
$Z'$ masses ought to be $\CO(1\,\tev)$. In the Nambu--Jona-Lasinio (NJL)
approximation---which we rely heavily upon here---the degree to which this
naturalness criterion is met is quantified by the ratio~\cite{cdt}
\be\label{eq:tune}
{\alpha(V_8) + \alpha(Z') - (\alpha^*(V_8) + \alpha^*(Z'))\over{
\alpha^*(V_8) + \alpha^*(Z')}} = {\alpha(V_8)\ts r_{V_8} + \alpha(Z')\ts
r_{Z'} \over {\alpha(V_8)(1-r_{V_8}) + \alpha(Z')(1-r_{Z'})}} \ts.
\ee
Here,
\bea\label{rdef}
\alpha(V_8) &=& {4\alpha_{V_8} \cos^4\thc\over{3\pi\ts}}, \quad
\alpha(Z') = {\alpha_{Z'} Y_{t_L} Y_{t_R} \cos^4\thy\over{\pi}} \ts;\nn\\
\tan\thc &=& {g_2\over{g_1}} \ts, \quad \tan\thy = {g'_2\over{g'_1}} \ts,
\quad
r_i = {\hat m^2_t\over{M^2_i}} \ts \ln \left({M^2_i
\over{\hat m^2_t}}\right) \ts, \hskip0.25truein (i = V_8, Z')\ts;
\eea
and $Y_{t_{L,R}}$ are the $\uone$ charges of $t_{L,R}$. The NJL condition on
the critical couplings for top condensation is $\alpha^*(V_8) + \alpha^*(Z')
= 1$. In this letter, we show that, for such large couplings, TC2 is tightly
constrained by the magnitude of $\ol B_d$--$B_d$ mixing: it requires $M_{V_8}
\simeq M_Z' \simge 5\,\tev$~\cite{others}. This implies that the topcolor
coupling $\alpha(V_8) + \alpha(Z')$ must be within less than 1\% of its
critical value, a tuning we regard as unnaturally fine.

There are two variants of TC2: The ``standard'' version~\cite{tctwohill}, in
which only the third generation quarks are $\suone$ triplets, and the
``flavor--universal'' version~\cite{ccs} in which all quarks are $\suone$
triplets. In standard TC2, $V_8$ and $Z'$ exchange gives rise to
flavor--changing neutral currents (FCNC) that mediate $\vert\Delta B\vert =
2$. In flavor--universal TC2, only $Z'$ exchange can generate such FCNC. Our
results constrain both types of TC2 theory.

The coloron interaction at energies well below $M_{V_8}$ is
\be\label{eq:HTCT}
\chv = {g^2_{V_8} \over{2 M^2_{V_8}}} \sum_{A=1}^8 J^{A\mu} J^A_\mu \ts,
\ee
where the coloron current written in terms of electroweak eigenstate (primed)
fields is given by
\be\label{eq:JVeight}
 J^A_\mu = \cos^2\thc \sum_{i=t,b} \ol q'_i \gamma_\mu \ts
 {\lambda_A\over{2}} \ts  q'_i - \sin^2\thc \sum_{i=u,d,c,s} \ol q'_i
 \gamma_\mu\ts  {\lambda_A\over{2}} \ts q'_i
   \ts.
\ee
The dominant coloron interactions for $\vert \Delta B\vert =2$ come from $\ol
b' b' \ol b' b'$ terms in Eq.~(\ref{eq:HTCT}). When written in terms of mass
eigenstate fields, they are (neglecting smaller terms proportional to
$\alpha_C$ alone):
\be\label{eq:HVeight}
\chv = {2\pi\alpha_C \cot^2\thc \over{M^2_{V_8}}}
\sum_{\lambda_1, \lambda_2 = L,R}
\biggl(D^*_{\lambda_1 bb} D_{\lambda_1 bd_i} \ts
D^*_{\lambda_2 bb} D_{\lambda_2 bd_i}
 \ts\ts \ol b_{\lambda_1} \gamma^\mu \ts
  {\lambda_A\over{2}} \ts d_{i\lambda_1}
\ts\ts \ol b_{\lambda_2} \gamma_\mu \ts {\lambda_A\over{2}} \ts
  d_{i\lambda_2} + {\rm h.c.} \biggr) 
\ee
Here, $d_i = d$ or $s$, and the unitary matrices $D_{L,R}$ arise from vacuum
alignment in the down--quark sector. We discuss them shortly.

In order that $Z'$ exchange not induce large $\vert \Delta S\vert = 2$
transitions, quarks of the two light (electroweak basis) generations must
have the same $\uone$ charge. Then, the $Z'$ interaction for $\vert\Delta B
\vert = 2$ is
\be\label{eq:HZprime}
\chz = {2\pi\alpha_Y \cot^2\thy \over{M^2_{Z'}}}
\sum_{\lambda_1, \lambda_2 = L,R}
\biggl(D^*_{\lambda_1 bb} D_{\lambda_1 bd_i} \ts
D^*_{\lambda_2 bb} D_{\lambda_2 bd_i} \ts \Delta Y_{\lambda_1}
\Delta Y_{\lambda_2}  \ts\ts \ol b_{\lambda_1} \gamma^\mu \ts d_{i\lambda_1}
\ts\ts \ol b_{\lambda_2} \gamma_\mu \ts d_{i\lambda_2} + {\rm h.c.} \biggr)
\ee
Here, $\Delta Y_{\lambda} = Y_{b\lambda} - Y_{d\lambda}$ is the difference of
$\uone$ charges.

Vacuum alignment in the technifermion sector leads to unitary matrices $W=
(W^U, W^D)$ which represent the mismatch between the directions of
spontaneous and explicit breaking of technifermion chiral symmetries. A
common feature of these matrices is that all their phases are rational
multiples of $\pi$. That is, for $N$ technifermion doublets, these phases may
be integral multiples of $\pi/N'$ for {\it various} $N'$ from~1
to~$N$~\cite{vacalign}. Extended technicolor couples quarks to technifermions
and generates the ``primordial'' quark mass matrices $\CM_{qij} =
\Lambda^{qT}_{iIJj} \ts W^T_{IJ} \ts \Delta_T$ (except for the $\hat m_t$
part of $\CM_{u\ts tt}$). Here $(q,T) = (u,U)$ or $(d,D)$; $i,j = u,c,t$ or
$d,s,b$; $I,J$ label the technifermion flavors. The $\Lambda^{qT}_{iIJj}$ are
{\it real} ETC couplings of order $(100-1000\,\tev)^{-2}$ and $\Delta_T$ is
the real technifermion condensate renormalized at the ETC breaking scale. If
the $\Lambda^{qT}_{iIJj}$ properly image $W^T$'s rational phases onto
$\CM_q$, there will be no strong CP violation, i.e., $\ol \theta_q =
\arg\det(\CM_u) + \arg\det(\CM_d) \simle \CO(10^{-10})$.  In this imaging,
all elements of $\CM_u$ and $\CM_d$ have (generally different) rational
phases that add to zero in the determinant. We assume this can happen in ETC
models.

Vacuum alignment in the quark sector~\cite{dashen} is achieved by minimizing
the quark energy $E_q(U,D) = -{\rm Tr}(\CM_u U^\dagger + \CM_d D^\dagger +
{\rm h.c.})$. The up and down--quark alignment matrices $U = U_L U_R^\dagger$
and $D = D_L D_R^\dagger$ are $3\times 3$ diagonal blocks in the $(SU(6)_L
\otimes SU(6)_R)/SU(6)_V$ matrices.\footnote{Actually, quark vacuum alignment
is based on first--order chiral perturbation theory, so it is inapplicable to
the heavy quarks $c,b,t$. When $\ol \theta_q = 0$, Dashen's procedure is
equivalent to making the mass matrices diagonal, real, and
positive~\cite{nuyts}. Thus, it correctly determines the quark unitary
matrices $U_{L,R}, D_{L,R}$ and the magnitude of strong and weak CP
violation.}  The ETC and TC2 interactions restrict the texture of the $\CM_u$
and $\CM_d$. This, in turn, determines the form of $U_{L,R}$ and $D_{L,R}$
and, ultimately, of the TC2 amplitude for $\ol B_d$--$B_d$ mixing.

Flavor--changing neutral current limits imply that ETC contributions to
$\CM_{u,d}$ are at most a few GeV, just enough to produce $m_b(M_{ETC})$. The
TC2 interactions generate $\hat m_t$, but off-diagonal elements in the third
row and column of $\CM_u$ come from ETC. They are expected to be no larger
than the 0.01--1.0~GeV associated with $m_u$ and $m_c$. Thus, $\CM_u$ is very
nearly block--diagonal and, so, $|U_{L,R \ts tu_i}| \cong |U_{L,R \ts u_it}|
\cong \delta_{tu_i}$. Limits on $\ol B_d$--$B_d$ mixing induced by exchange
of ``bottom pions''~\cite{kominis,bbhk}, together with the need to generate
appropriate intergenerational mixing in the Cabibbo--Kobayashi--Maskawa (CKM)
matrix $V = U^\dagger_L D_L$, require that $d_R,s_R \leftrightarrow b_L$
elements of $\CM_d$ are much smaller than the $d_L,s_L \leftrightarrow b_R$
elements~\cite{tctwoklee}. This makes $D_R$ nearly $2\times 2$ times $1\times
1$ block--diagonal.

Since $D_L$ contains almost all the mixing between $b$ and $d,s$ and, hence,
between the third generation and the first two, the mixing pattern of $V =
U^\dagger_L D_L$ is essentially the same as that of $D_L$. To an excellent
approximation,
\bea\label{eq:ckm}
\vert V_{td_i} \vert &=& \vert U^*_{L tt} D_{L bd_i} \vert =
 \vert D_{L bd_i}\vert \ts; \nn\\
V^*_{tb} V_{td_i} &=& U_{L tt} D^*_{L bb}  U^*_{L tt} D_{L bd_i}
= D^*_{L bb} D_{L bd_i} \ts.
\eea
Our strongest limit on the $V_8$ and $Z'$ masses will arise from
Eq.~(\ref{eq:ckm}).

The dominant TC2 contribution to $|\Delta B_d|=2$ is given by the LL terms in
$\chv + \chz$. The relevant observable is the $B^0_L$--$B^0_S$ mass
difference. Since $\Gamma_{12} \ll M_{12}$, it is given by $\Delta M_{B_d} =
2|M_{12}|$~\cite{buras}. Fierzing $\chv$ into a product of color--singlet
currents and calculating the $\ol B_d$--$B_d$ matrix element in the usual
vacuum--insertion approximation, the TC2 contribution to $M_{12}$ is
\be\label{eq:DelMbd}
2 (M_{12})_{\rm TC2} 
={4\pi\over{3}}\left[{\alpha_C \cot^2\thc\over{3 M^2_{V_8}}} + 
{\alpha_Y \cot^2\thy (\Delta Y_L)^2 \over{M^2_{Z'}}} \right]
\ts \eta_B M_{B_d} f^2_{B_d} B_{B_d} (D^*_{Lbb} D_{Lbd})^2
\ts. 
\ee
Here, $\eta_B = 0.55 \pm 0.01$ is a QCD radiative correction factor for the
LL product of color--singlet currents. We take $f_{B_d} \sqrt{B_{B_d}} =
(200\pm 40)\mev$~\cite{buras}, where $f_{B_d}$ and $B_{B_d}$ are,
respectively, the $B_d$--meson decay constant and bag parameter. This TC2
contribution is to be added to the standard model one,
\be\label{eq:DelMSM}
2 (M_{12})_{\rm SM} = {G^2_F\over{6\pi^2}} \eta_B M_{B_d} f^2_{B_d}
M^2_W S_0(x_t) (V^*_{tb} V_{td})^2 \ts,
\ee
where the top--quark loop function $S_0(x_t) \cong 2.3$ for $x_t =
m_t^2(m_t)/M^2_W$ and $m_t(m_t) = 167\,\gev$. 

\begin{figure}[h]
\leavevmode
\centering
\epsfig{file=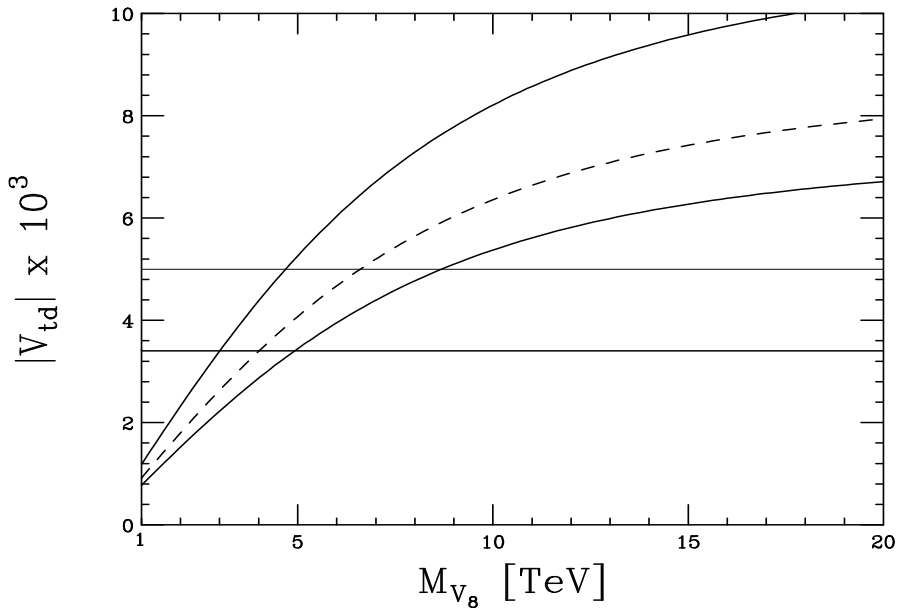,width=8cm,height=11.8cm,angle=90}
\caption{The allowed region in the $(M_{V_8},|V_{td}|)$ plane. 
The upper horizontal line comes from the lower limit on $|V_{td}|$ derived by
making use of $\epsilon$ plus the CKM unitarity triangle. The lower line is
the lowest value of $|V_{td}|$ for which ${\rm Im}(V_{ts}^*V_{td})$ and,
therefore, $\epsilon'/\epsilon$ are not too small.}
\label{bmixI}
\end{figure}

To determine the $\Delta M_{B_d}$--restriction on TC2, we adopt the
ETC--based analysis of quark mixing matrices made above. Then, $D^*_{Lbb}
D_{Lbd} = V^*_{tb} V_{td} \cong V_{td}$, so that the standard model and TC2
contributions add coherently. Next, to simplify our discussion, we ignore for
now the $Z'$ contribution to $\Delta M_{B_d}$ and $\hat m_t$. The NJL
approximation then implies $\cot^2\thc = 3\pi/4\alpha_C(1\,\tev) \simge 25$
for $\alpha(V_8) \ge \alpha^*(V_8) = 1$ and $\alpha_C(1\,\tev) \simeq
0.093$. Using $\Delta M_{B_d} = (3.11\pm 0.11) \times
10^{-13}\,\gev$~\cite{pdg}, we plot in Fig.~\ref{bmixI} the constraint in the
$(M_{V_8}, \vert V_{td}\vert)$ plane. The width of the allowed band comes
mostly from the error in $f_{B_d}\sqrt{B_{B_d}}$.

To set a bound on $M_{V_8}$, we must find the lowest possible value of
$|V_{td}|$. If ETC contributions to the kaon CP--violating parameter
$\epsilon$ are negligible, this lowest $|V_{td}|$ can be read off the
$(\bar\rho,\bar\eta)$ plane fits using $\epsilon$ and
$|V_{ub}/V_{cb}|$. (Data on $\Delta M_{B_{d,s}}$ are not used as these
quantities are affected by TC2.) The most recent fit~\cite{caravaglios} gives
$\bar\rho < 0.40$ and $\bar\eta > 0.20$ at the 95\% C.L. This yields
\be\label{eq:Vtdhigh} 
|V_{td}|=\lambda|V_{cb}|\,\sqrt{(1-\bar\rho)^2 + \bar\eta^2} > 5\times
10^{-3} \ts,
\ee
where we used the $2\sigma$ lower limit value $|V_{cb}|=0.036$. This bound is
displayed as the upper horizontal line in Fig.~\ref{bmixI}. Its intersection
with the left edge of the band gives the conservative lower limit for
$M_{V_8}$,
\be\label{eq:MVlimita}
M_{V_8} \ge 4.8\,\tev \quad \Ra \quad
{\alpha(V_8) - \alpha^*(V_8)\over{\alpha^*(V_8)}} \le 0.0075 \ts,
\ee
The tuning of $\alpha(V_8)$ is unnaturally fine.

If ETC contributes significantly to $\epsilon$, we must remove its
constraint from the standard model $(\ol \rho,\ol \eta)$ plane. The remaining
limitation on $|V_{td}|$, other than unitarity of $V$, comes from
$\epsilon'/\epsilon$. We expect that $\epsilon'$ is unaffected by
ETC~\cite{blt}. In the standard--model Wolfenstein
parameterization~\cite{pdg}, it is given by
\be\label{eq:epsprime}
{\epsilon'\over{\epsilon}} \equiv {\rm Im}(V^*_{ts} V_{td})S = A^2 \lambda^5
\ol \eta S \ts.
\ee
Here, $A = |V_{cb}|/\lambda^2=0.83$ is obtained using $|V_{cb}|=0.04$.  Also,
$S = P^{(1/2)}-P^{(3/2)}$ where $P^{(\Delta I)}$ contains the hadronic matrix
elements in the $|\Delta I| = 1/2$ and 3/2 amplitudes~\cite{bosch}. In
Eq.~(\ref{eq:epsprime}), $\epsilon$ is taken from experiment, so that
potential ETC contributions to it are not an issue. From Ref.~\cite{bosch},
hadronic matrix element calculations imply $S < 21.6$. The world average is
$\epsilon'/\epsilon = (19.3 \pm 2.4)\times 10^{-4}$~\cite{epexp}. The
experimental situation is still somewhat unsettled; we conservatively assume
the lower limit $\epsilon'/\epsilon > 12\times 10^{-4}.$ This gives $\ol \eta
> 0.156$. Using this, and the 95\% C.L. upper limit $|V_{ub}/V_{cb}| < 0.14$,
we find $\ol \rho < 0.60$ and
\be\label{eq;Vtdlow}
|V_{td}| > 3.4\times 10^{-3} \ts.
\ee
Intersecting this value with the allowed band in Fig.~\ref{bmixI} yields
\be\label{eq:MVlimitb}
M_{V_8} \ge 3.1\,\tev \quad \Ra \quad
{\alpha(V_8) - \alpha^*(V_8)\over{\alpha^*(V_8)}} \le 0.016
\ee
Although somewhat a matter of taste, we regard this level of fine tuning at
the edge of acceptability for a dynamical theory with naturalness as a design
goal.

To estimate the effect of the $Z'$ on this naturalness criterion, we suppose
that $V_8$ and $Z'$ exchange contribute equally to generating $\hat m_t$,
i.e., that $\alpha^*(V_8) = \alpha^*(Z') = \half$ in the NJL
approximation. We also assume $(\Delta Y_L)^2 \simeq Y_{t_L} Y_{t_R}$. Then,
$\cot^2\thc \ge \cot^2\thc^* = 3\pi/8\alpha_C(1\,\tev) = 12.5$ and
\be\label{eq:VZ}
{\alpha_C \cot^2\thc \over{3 M^2_{V_8}}} + {\alpha_Y \cot^2\thy
(\Delta Y_L)^2  \over{M^2_{Z'}}} \simge {\pi\over{8}}
\left({1\over{M^2_{V_8}}} + {4\over{M^2_{Z'}}}\right) \ts.
\ee
Equating the right--hand side to its maximum value $\pi/(4(4.8\,\tev)^2)$
obtained when we neglected $Z'$ gives
\be\label{eq:MZlimit}
M_{Z'} = 2\left[{2\over{(4.8\,\tev)^2}} -
  {1\over{M^2_{V_8}}}\right]^{-\half} > 6.8\,\tev
\quad \Ra \quad {\alpha(Z') - \alpha^*(Z')\over{\alpha^*(Z')}} < 0.0042
\ee when $M_{V_8} \ra \infty$. For $M_{V_8} = 4.8\,\tev$, we have $M_{Z'} =
9.6\,\tev$ and
\be\label{eq:VZtune}
{\alpha(V_8)\ts r_{V_8} + \alpha(Z')\ts r_{Z'} \over
  {\alpha(V_8)(1-r_{V_8}) + \alpha(Z')(1-r_{Z'})}}  \simeq 0.0049 \ts.
\ee
In short, adding the $Z'$ contribution to $\Delta M_{B_d}$ does not make TC2
more natural.

It is possible to avoid this naturalness problem if, contrary to ETC
expectations, all mixing between the third generation and the two light ones
is contained in the up--sector matrices $U_{L,R}$. Even this possibility is
constrained by $V_8$ and $Z'$--exchange contributions to the neutron electric
dipole moment, $d_n$. They affect only the up--quark moment. The $V_8$
contribution suffices; it is
\be\label{eq:upmom}
(d_u)_{\rm TC2} = {2e\over{3}}\ts
\frac{\alpha_C \cot^2\theta_C}{4\pi}\;\frac{4}{3}\,\frac{m_t(m_t)}{M_{V_8}^2}\,
\,{\rm Im}(U^*_{Ltu} U_{Ltt} U^*_{Rtt} U_{Rtu}) \ts.
\ee
The limit~\cite{pdg} $d_n < 0.63\times 10^{-25}\,e$--cm
  yields~\footnote{${\rm Im}(U^*_{Ltu} U_{Ltt} U^*_{Rtt} U_{Rtu})$ = 0 in ETC,
  but we cannot appeal to that here.}
\be\label{eq:edmlimit} 
M_{V_8} \simge 1\,\tev \ts\ts   \sqrt{{|{\rm Im}(U^*_{Ltu} U_{Ltt} U^*_{Rtt}
    U_{Rtu})|\over{10^{-7}}}} \ts.
\ee
Since $U_{Ltu} = \sum_{d_i} D_{Lbd_i} V^*_{ud_i} \cong D_{Lbb} V^*_{ub}$
implies $|U_{Ltu}| \simge 2\times 10^{-3}$, a natural $V_8$ mass of a
1--2~TeV suggests $|U_{Rtu}| \simle 5\times 10^{-5}$. Thus, the analog of the
Kominis constraint~\cite{kominis,bbhk} applies to $U_R$.

In summary, generation of light quark masses via extended technicolor
strongly suggests that all quark mixing occurs in the left--handed down
sector. Then, the TC2 mechanism for $m_t$ requires, {\it in the NJL
approximation}, $V_8$ and $Z'$ masses exceeding $5\,\tev$. This, in turn,
needs fine--tuning of the topcolor couplings to within less than 1\% of their
critical values. We do not know how this difficulty will be resolved.

{\it Note added in proof:} The contribution to $Z^0 \ra b \ol b$ from
top--pions, the pseudoGoldstone bosons arising from top condensation, also
significantly constrains TC2~\cite{bk}. This contribution can be made
consistent with experiment by increasing the top--pion decay constant and/or
its mass. Increasing either requires raising the TC2 scale embodied
in $M_{V_8}$ and $M_{Z'}$. In the NJL approximation, we estimate that this
implies a fine--tuning of about 1\%, comparable to what we found above from
$\Delta M_{B_d}$.

We have benefitted from conversations with S.~Chivukula, L.~Giusti, C.~Hill,
and E.~Simmons. G.B. thanks Lawrence Berkeley National Laboratory for
its hospitality during the latter stages of this work. This research was
supported in part by the U.~S.~Department of Energy under
Grant~No.~DE--FG02--91ER40676.

\bigskip
  %

%

\vfil\eject

\end{document}